\documentclass[aps,pre,
tightenlines,
preprint,
showpacs,amsfonts,amsmath,amssymb
]{revtex4}
\usepackage{graphicx}

\begin{document}

\title{Generalized theory for numerical instability of the Gaussian-filtered 
Navier-Stokes equations as a model system for large eddy simulation of 
turbulence}

\author{Masato Ida}\email[E-Mail: ]{ida.masato@jaea.go.jp}
\affiliation{Center of Computational Science and E-systems, Japan Atomic Energy Agency, 6-9-3 Higashi-Ueno, Taito-ku, Tokyo 110-0015, Japan}

\author{Nobuyuki Oshima}
\affiliation{Division of Mechanical and Space Engineering, Graduate School of Engineering, Hokkaido University, Kita 13 Nishi 8, Kita-ku, Sapporo 060-8628, Japan}

\begin{abstract}
The Gaussian-filtered Navier-Stokes equations are examined theoretically and 
a generalized theory of their numerical stability is proposed. Using the 
exact expansion series of subfilter-scale stresses or integration by parts, 
the terms describing the interaction between the mean and fluctuation 
portions in a statistically steady state are theoretically rewritten into a 
closed form in terms of the known filtered quantities. This process involves 
high-order derivatives with time-independent coefficients. Detailed 
stability analyses of the closed formulas are presented for determining 
whether a filtered system is numerically stable when finite difference 
schemes or others are used to solve it. It is shown that by the Gaussian 
filtering operation, second and higher even-order derivatives are derived 
that always exhibit numerical instability in a fixed range of directions; 
hence, if the filter widths are unsuitably large, the filtered Navier-Stokes 
equations can in certain cases be unconditionally unstable even though there 
is no error in modeling the subfilter-scale stress terms. As is proved by a 
simple example, the essence of the present discussion can be applied to any 
other smooth filters; that is, such a numerical instability problem can 
arise whenever the dependent variables are smoothed out by a filter.
\end{abstract}


\maketitle

\section{Introduction}
\label{secI}
Large eddy simulation (LES) \cite{ref1,ref2} is one of the typical numerical 
approaches to turbulent flows, in which large-scale structures in fluid 
motion are solved directly while the effects of the small-scale eddies are 
modeled based on a filtering operation that separates the high- and 
low-wavenumber modes in turbulent flow fields. Because LES enables us to 
treat time-dependent, high-Reynolds-number turbulence with substantially 
smaller computational effort and storage than direct numerical simulation 
(DNS) resolving all scales of motion, it has been used in many applications 
in a variety of research fields, including fluid machinery \cite{ref3}, combustion 
engineering \cite{ref4,ref5}, atmospheric science \cite{ref6,ref7}, geophysics \cite{ref8,ref9,ref10}, and 
astrophysics \cite{ref11,ref12}.

For the past few years, however, several reports have been published that 
point out the incompleteness of current LES modeling. In those repots it has 
been implied that even a completely accurate LES model could be numerically 
unstable. In Ref.~\cite{ref13}, Leonard showed that the tensor-diffusivity model, 
re-derived by truncating an exact expansion series of subfilter-scale forces 
\cite{ref14,ref15,ref16}, works as a negative-diffusion term in the stretching 
directions of fluid motion and hence, that it could lead to numerical 
instability when used for finite difference schemes. Since that model is 
exact for first-order velocity fields where the velocity components can be 
described by linear polynomials, the model's negative diffusivity can be 
considered to exhibit the nature of the {\it exact} model under a particular condition. 
Also, Winckelmans et al. \cite{ref20} and Kobayashi and Shimomura \cite{ref17,ref17a,ref17b} 
pointed out that the tensor-diffusivity model behaves unstably in a plane 
channel flow, also due to the model's negative diffusivity. In a comparative 
study of LES models (the tensor-diffusivity and rational LES models) \cite{ref36}, 
Iliescu et al. showed that the tensor-diffusivity model can be unstable in a 
high-Reynolds-number driven cavity flow, as well. In Ref.~\cite{ref18}, Ida and 
Taniguchi derived a closed form of the Gaussian-filtered Navier-Stokes 
(GFNS) equations under a simple assumption about the instantaneous velocity 
profile and showed theoretically that the shears in time-averaged flow 
fields can also be the seed of the numerical instability of the filtered 
system, because a cross derivative of the filtered velocity component, being 
unconditionally unstable in numerical simulation, appears in the closed 
formula. The authors stressed that the unstable portions in the closed 
formula must be solved accurately without using artificial techniques (e.g., 
clipping or damping), since those portions derive naturally from the 
filtering operation and are thus a part of the governing equations for LES. 
In the sequel to that paper \cite{ref19}, Ida and Taniguchi further ascertained that 
the shears in the time-averaged fields can, through the filtering operation, 
cause the appearance of a numerically unstable term that always exhibits a 
negative diffusivity in a fixed direction, a conclusion that is able to 
explain the problematic instability that has frequently been confronted in 
wall-bounded turbulent flow computations (e.g., Ref.~\cite{ref20,ref17}), where a 
strong shear appears in the time-averaged streamwise velocity. The 
theoretical and numerical findings listed above appear to suggest that the 
filtering operation itself is the underlying cause of the numerical 
instability in LES, raising the question whether a numerically stable LES 
model can be ideally accurate or not.

A similar scenario can be found in simulation strategies for collisionless 
plasma kinetics that use the Vlasov-Poisson or Vlasov-Maxwell system as a 
governing equation. In Refs.~\cite{ref21,ref22}, Klimas has attempted to apply a 
Gaussian filter to the Vlasov equations in order to mollify the 
filamentation of the distribution function (an infinitely fine structure in 
the phase space), and found that the filtered Vlasov equations can be 
rewritten into a closed form in terms of the filtered distribution function 
and are thus solvable without any empirical modeling. (We note here that in 
Klimas's study the filtering operation was only applied in the velocity 
space, which allows for relatively easy derivation of closed formulas and 
results in only a few additional terms.) In that closed formula, a cross 
derivative of the filtered distribution function appears, which, as Figua et 
al. suggested \cite{ref23} (see also Refs.~\cite{ref17,ref18}), makes the filtered system 
ill-conditioned and unsuitable for numerical simulations using finite 
difference methods or others excluding the spectral method. As with the 
Navier-Stokes cases mentioned above, that finding implies that the Gaussian 
filtering operation itself, and not the modeling or approximations, 
destabilizes the governing equations.

The present paper extends the numerical stability analysis of the GFNS 
equations performed by Ida and Taniguchi \cite{ref19} to construct a generalized 
theory for the numerical instability of the system. The present discussion 
assumes that the flow fields are, as in Ref.~\cite{ref19}, in a statistically steady 
state (an assumption allowing us to decompose the velocity components into 
time-independent mean and time-dependent fluctuation portions), but the mean 
velocities may be described by high-order polynomials in terms of the 
spatial coordinates, while in Ref.~\cite{ref19} first-order velocity fields have 
mainly been considered. Under these assumptions and using the exact 
expansion series or integration by parts, we rewrite the terms that 
represent the interaction between the mean and fluctuation portions 
(referred to below as ``mean-fluctuation terms''), filtered by a Gaussian 
function, into closed forms involving high-order cross derivatives, and show 
that through Gaussian filtering, various kinds of unconditionally unstable 
terms having time-independent coefficients are derived which numerically 
destabilize the modes in a fixed range of directions. Also, detailed 
stability analyses of the resulting closed formulas are presented to derive 
a stability criterion for the choice of filter widths. In the present paper, 
for simplicity we only discuss cases in which the mean velocity field has 
one-dimensional (1D) or two-dimensional (2D) structures. Moreover, we are 
not concerned with the commutation error between differentiation and 
filtering (see e.g. Refs.~\cite{ref24,ref25,ref26,ref27} for recent efforts to resolve the 
commutation error), assuming each filter width to be constant in the 
corresponding spatial direction. This treatment warrants that the 
numerically unstable terms that we discuss are not those originating from 
the commutation error, which is a modeling failure.

The present theoretical investigation has been performed assuming the use of 
finite difference schemes. However, most of the results will be true for 
other numerical methods (e.g., finite volume, finite element, compact 
differencing) as well. Also, in order to accomplish the theoretical 
investigation without the aid of numerical analysis, the present study 
neglects the cutoff of high-wavenumber modes originating from the use of finite grid spacing. 
The Gaussian filter 
considered in the present study is, therefore, assumed to approximately 
represent the numerical damping of high-wavenumber modes due to numerical 
viscosity (also originating from the use of finite grid spacing), and also to 
be an explicit filter applied independently of numerical 
discretization \cite{ref28,ref20}. Because of these assumptions, we use the 
term ``subfilter scale'' in stead of ``subgrid scale'' throughout this paper.

The present paper is organized as follows. In Sec.~\ref{secII}, the governing 
equations and definitions useful for the present investigation are 
introduced. In Sec.~\ref{secIII}, the numerical stability of arbitrary-order partial 
differential equations involving high-order cross derivatives is 
theoretically discussed to derive a stability criterion for them, which is 
essential for our study. Combining the result of this stability analysis 
with an exact expansion series or integration by parts allows us to 
construct a generalized theory for the filtering instability under 
statistically steady-state conditions. In Sec.~\ref{secIV}, several specific examples 
are investigated to elucidate how the stability criterion restricts the 
choice of filter widths, and in Sec.~\ref{secV}, to elucidate a fundamental mechanism 
of the numerically unstable terms and to show that the essential part of the 
present results is true for non-Gaussian smooth filters as well, a simple 
advection problem is considered where the true solution has a discontinuous 
step. Section \ref{secVI} presents notes on remaining issues that must be resolved to 
gain a more generalized theory. As stated in that section, the present 
theory has several limitations in its applicability. We conceive of the 
present theory as an intermediate step towards a complete theory of the 
numerical instability of the GFNS equations. Section \ref{secVII} summarizes this 
paper, and the appendix presents a mathematical proof of the exact expansion 
series using elementary mathematics, thereby assuring the self-consistency 
of the present paper.

\section{Governing equations, filtering operations, and definitions}
\label{secII}
Incompressible viscous fluid flows are described by the Navier-Stokes 
equations:
\begin{equation}
\label{eq1}
\frac{\partial u_i }{\partial t}+\frac{\partial u_j u_i }{\partial x_j 
}=-\frac{\partial p}{\partial x_i }+\nu \frac{\partial ^2u_i }{\partial x_j 
\partial x_j }\quad \mbox{for }i=1,2,3,
\end{equation}
\begin{equation}
\label{eq2}
\frac{\partial u_i }{\partial x_i }=0,
\end{equation}
where the summation convention is assumed, and $u_i $ ($i=1,2,3)$ are the 
velocity components, $p$ is the pressure divided by the constant fluid 
density, and $\nu $ is the kinematic viscosity. The subfilter-scale terms, 
resulting from a low-pass filtering operation, are derived from the 
convection terms (i.e., the second term of Eq.~(\ref{eq1})). In what follows, we 
assume that the velocity components can be decomposed into time-averaged and 
fluctuation portions as
\begin{equation}
\label{eq3}
{\rm {\bf u}}({\rm {\bf x}},t)={\rm {\bf U}}({\rm {\bf x}})+{\rm {\bf 
u'}}({\rm {\bf x}},t),
\end{equation}
where ${\rm {\bf u}}=(u_1 ,u_2 ,u_3 )$, ${\rm {\bf U}}=(U_1 ,U_2 ,U_3 )$, 
and ${\rm {\bf u'}}=(u'_1 ,u'_2 ,u'_3 )$. We also assume that the 
mean velocity ${\rm {\bf U}}$ is time-independent; i.e., that the flow is in 
a statistically steady state. From Eqs.~(\ref{eq2}) and (\ref{eq3}), one can derive
\begin{equation}
\label{eq4}
\frac{\partial U_i }{\partial x_i }=\frac{\partial u'_i }{\partial x_i 
}=0.
\end{equation}
Using Eqs.~(\ref{eq3}) and (\ref{eq4}), the convection term in Eq.~(\ref{eq1}) is rewritten as 
follows:
\begin{equation}
\label{eq5}
\frac{\partial u_j u_i }{\partial x_j }=h_i +u'_j \frac{\partial u'_i 
}{\partial x_j }+U_j \frac{\partial U_i }{\partial x_j },
\end{equation}
\begin{equation}
\label{eq6}
h_i \equiv U_j \frac{\partial u'_i }{\partial x_j }+\frac{\partial U_i 
}{\partial x_j }u'_j \quad \mbox{for }i=1,2,3,
\end{equation}
where $h_i $ represents the interaction between the mean and fluctuation 
portions, on which we focus our attention.

The filter function is assumed to be Gaussian:
\[
{\mathcal G}(X;\Delta )=\sqrt {\frac{\gamma }{\pi \Delta ^2}} \exp \left( 
{-\frac{\gamma X^2}{\Delta ^2}} \right),
\]
which satisfies $\int_{X=-\infty }^{X=\infty } {{\mathcal G}(X;\Delta )} dX=1$, where 
$\gamma $ is commonly set to $6$ in LES and $\Delta $ is the filter width. 
Using this, the filtering operation in the $x_i $ direction is performed as 
a convolution integral:
\[
\begin{array}{c}
 \bar {f}(x_i ,\ldots ,t)=\int\limits_{X=-\infty }^{X=\infty } {{\mathcal G}(x_i 
-X;\Delta _i )f(X,\ldots ,t)dX} \\ 
 \equiv {\mathcal G}_i \star f, \\ 
 \end{array}
\]
where the overbar denotes the filtered quantities. Three-dimensional (3D) 
filtering is achieved by successively performing this convolution as 
follows:
\[
\begin{array}{c}
 \bar {f}({\rm {\bf x}},t)=\int\limits_{{\rm {\bf X}}=-\infty }^{{\rm {\bf 
X}}=\infty } {\prod\limits_{i=1}^3 {{\mathcal G}(x_i -X_i ;\Delta _i )} f({\rm {\bf 
X}},t)d{\rm {\bf X}}} ={\mathcal G}_1 \star [{\mathcal G}_2 \star ({\mathcal G}_3 \star f)] \\ 
 \equiv {\mathcal G}_{123} \star f, \\ 
 \end{array}
\]
where $\Delta _i $ ($i=1,2,3)$ is the filter width in the $x_i $ direction. 
As stated in Sec.~\ref{secI}, we assume throughout this paper that each filter width 
is a constant, and thus
\[
{\mathcal G}_i \star \frac{\partial f}{\partial x_j }=\frac{\partial ({\mathcal G}_i \star 
f)}{\partial x_j },
\]
\[
{\mathcal G}_i \star ({\mathcal G}_j \star f)={\mathcal G}_j \star ({\mathcal G}_i \star f)\quad \mbox{for }i,j=1,2,3.
\]
For the convenience of the following discussion, we introduce the {\it residual stress function} defined 
as
\begin{equation}
\label{eq7}
{\mathcal R}_a [F(U_1 ,u'_1 ,\ldots )]\equiv {\mathcal G}_a \star F(U_1 ,u'_1 ,\ldots )-F({\mathcal G}_a 
\star U_1 ,{\mathcal G}_a \star u'_1 ,\ldots ),
\end{equation}
which yields, for example,
\begin{equation}
\label{eq8}
{\mathcal R}_1 \left[ {U_j \frac{\partial u'_i }{\partial x_j }} \right]={\mathcal G}_1 \star 
\left( {U_j \frac{\partial u'_i }{\partial x_j }} \right)-({\mathcal G}_1 \star U_j 
)\frac{\partial ({\mathcal G}_1 \star u'_i )}{\partial x_j },
\end{equation}
\begin{equation}
\label{eq9}
{\mathcal R}_{123} \left[ {u'_1 u'_2 } \right]={\mathcal G}_{123} \star (u'_1 u'_2 
)-({\mathcal G}_{123} \star u'_1 )({\mathcal G}_{123} \star u'_2 ).
\end{equation}

Based on the above assumptions and definitions, the Navier-Stokes equations 
filtered using a 3D Gaussian filter are written as
\begin{equation}
\label{eq10}
\frac{\partial \bar u'_i }{\partial t}+\frac{\partial \bar {u}_j \bar 
{u}_i }{\partial x_j }=-\frac{\partial \bar {p}}{\partial x_i }+\nu 
\frac{\partial ^2\bar {u}_i }{\partial x_j \partial x_j }-{\mathcal R}_{123} \left[ 
{\frac{\partial u_j u_i }{\partial x_j }} \right],
\end{equation}
or
\begin{eqnarray}
\label{eq11}
\frac{\partial \bar u'_i }{\partial t}+\frac{\partial \bar {u}_j \bar 
{u}_i }{\partial x_j } &=& -\frac{\partial \bar {p}}{\partial x_i }+\nu 
\frac{\partial ^2{\bar {u}}'_i }{\partial x_j \partial x_j }-{\mathcal R}_{123} \left[ 
{h_i } \right] \nonumber \\ 
 &-& {\mathcal R}_{123} \left[ {u'_j \frac{\partial u'_i }{\partial 
x_j }} \right]+\left( {\nu \frac{\partial ^2\bar {U}_i }{\partial x_j 
\partial x_j }-{\mathcal R}_{123} \left[ {U_j \frac{\partial U_i }{\partial x_j }} 
\right]} \right),
\end{eqnarray}
which can be considered an equation for both $\bar u'_i $ and $\bar 
{u}_i $ because $\partial \bar u'_i /\partial t=\partial \bar {u}_i 
/\partial t$. The terms in the last parentheses of Eq.~(\ref{eq11}) can be 
considered time-independent source terms, which may have no influence on the 
numerical stability. The next to last term represents the residual stress 
forces due to the nonlinear interaction between the fluctuation portions, 
the stability analysis of which is difficult to complete theoretically and 
thus requires numerical experiments. Although it has been pointed out that 
terms having the same form as ${\mathcal R}_{123} [u'_j \partial u'_i /\partial x_j 
]$ can instantaneously be numerically unstable \cite{ref13,ref18}, such terms should 
not necessarily lead to numerical instability in actual computation, because 
their time-averaged nature can be dissipative. In what follows, we only 
consider the numerical stability of
\begin{equation}
\label{eq12}
\nu \frac{\partial ^2{\bar {u}}'_i }{\partial x_j \partial x_j }-{\mathcal R}_{123} 
\left[ {h_i } \right],
\end{equation}
i.e., the difference between the molecular viscosity and the residual stress 
forces due to the mean-fluctuation interaction, and do not take into 
consideration the numerical effects of the nonlinear term. In this respect, 
our theoretical investigation is incomplete. Comments on potential 
approaches to resolving this incompleteness are given in Sec.~\ref{secV}. As shown in 
what follows, the closed form of Eq.~(\ref{eq12}) has time-independent coefficients, 
meaning that the numerical stability of this portion is time-independent.

We introduce here some mathematical tools that allow us to rewrite the 
filtered mean-fluctuation term ${\mathcal R}_{123} \left[ {h_i } \right]$ into a closed 
form. Yeo \cite{ref14} and others \cite{ref13,ref15,ref16} have derived a very interesting 
identity, which is applicable to all differentiable and continuous functions 
$f(x)$ and $g(x)$,
\begin{equation}
\label{eq13}
\overline {(fg)} -\bar {f}\bar {g}=\sum\limits_{n=1}^\infty 
{\frac{1}{n!}\left( {\frac{\Delta ^2}{2\gamma }} \right)^n\frac{\partial 
^n\bar {f}}{\partial x^n}\frac{\partial ^n\bar {g}}{\partial x^n}} .
\end{equation}
Here, the overbar indicates $x$-directional Gaussian filtering. It is worth 
noting that the right-hand side of this identity only involves the known 
filtered quantities $\bar {f}$ and $\bar {g}$. This outstanding feature of 
the series allows us to rewrite the residual ${\mathcal R}[fg]=\overline {(fg)} -\bar 
{f}\bar {g}$ into a closed form. For 2D Gaussian filtering in the $(x_1 ,x_2 
)$ plane, this series becomes
\begin{eqnarray}
\label{eq14}
\overline {(fg)} -\bar {f}\bar {g}=\sum\limits_{k=1}^2 {\left( 
{\frac{\Delta _k ^2}{2\gamma }} \right)\frac{\partial \bar {f}}{\partial x_k 
}\frac{\partial \bar {g}}{\partial x_k }} +\sum\limits_{k,l=1}^2 
{\frac{1}{2!}\left( {\frac{\Delta _k ^2}{2\gamma }} \right)\left( 
{\frac{\Delta _l ^2}{2\gamma }} \right)\frac{\partial ^2\bar {f}}{\partial 
x_k \partial x_l }\frac{\partial ^2\bar {g}}{\partial x_k \partial x_l }} \nonumber \\
 +\sum\limits_{k,l,m=1}^2 {\frac{1}{3!}\left( {\frac{\Delta _k ^2}{2\gamma 
}} \right)\left( {\frac{\Delta _l ^2}{2\gamma }} \right)\left( {\frac{\Delta 
_m ^2}{2\gamma }} \right)\frac{\partial ^3\bar {f}}{\partial x_k \partial 
x_l \partial x_m }\frac{\partial ^3\bar {g}}{\partial x_k \partial x_l 
\partial x_m }} +\cdots . 
\end{eqnarray}
The following identity is also useful:
\begin{equation}
\label{eq15}
{\mathcal G}_i \star (x_i f)=\left( {x_i +\frac{\Delta _i ^2}{2\gamma }\frac{\partial 
}{\partial x_i }} \right)({\mathcal G}_i \star f),\quad i=1,2,3,
\end{equation}
which can be derived using integration by parts (e.g., Refs.~\cite{ref21,ref23,ref31,ref18}). This yields, for example,
\begin{equation}
\label{eq16}
{\mathcal G}_i \star x_i =x_i ,
\end{equation}
\begin{equation}
\label{eq17}
{\mathcal G}_i \star x_i ^2=x_i ^2+\frac{\Delta _i ^2}{2\gamma },
\end{equation}
\begin{equation}
\label{eq18}
{\mathcal G}_i \star x_i ^3=x_i ^3+\frac{3\Delta _i ^2}{2\gamma }x_i ,
\end{equation}
where the summation convention is not adopted. Using Eq.~(\ref{eq16}), Eq.~(\ref{eq15}) can 
be rewritten into
\begin{equation}
\label{eq19}
{\mathcal R}_i [x_i f]=\frac{\Delta _i ^2}{2\gamma }\frac{\partial ({\mathcal G}_i \star 
f)}{\partial x_i },\quad i=1,2,3.
\end{equation}

The expansion series (\ref{eq13}) and (\ref{eq14}) enable us to derive a closed 
form of Eq.~(\ref{eq12}). As can be seen from these series, the closed form has 
high-order 
derivatives including high-order cross derivatives, and the coefficients of 
these derivatives are time-independent, such as $U_j $ and $\partial U_i 
/\partial x_j $ in $h_i $. The numerical stability of such high-order 
derivatives are examined below.

\section{Numerical stability of arbitrary-order partial differential 
equations}
\label{secIII}
We derive and examine in this section a numerical-stability criterion to 
determine whether an arbitrary-order partial differential equation (PDE) can 
be solved stably (and accurately) by a finite difference scheme. Although 
the numerical stability of PDEs is known to depend on the discretization 
scheme applied, we do not discuss a certain form of finite differencing. We 
instead consider the {\it exact} amplification factor of the PDEs, which is essential 
and may be sufficient for our aim. It is well known that a diffusion 
equation, for example, is numerically stable (numerically unstable) when the 
coefficient of the diffusion term is positive (negative), i.e., when the 
exact amplification factor is less than (greater than) unity. We assume here 
that such is also the case for other types of PDEs including high-order 
ones, and use their amplification factors to judge whether a stable finite 
difference scheme can exist for the corresponding PDE.

Let us consider the exact solution of a 2D arbitrary-order PDE,
\begin{equation}
\label{eq20}
\frac{\partial f}{\partial t}=\mu \left( {\frac{\partial }{\partial x}} 
\right)^m\left( {\frac{\partial }{\partial y}} \right)^nf,
\end{equation}
\[
f=f(x,y,t)
\]
where $\mu $ is a real constant and $m,n\ge 0$ are integers. The initial 
condition is
\begin{equation}
\label{eq21}
f(x,y,0)=\exp (\mbox{i}(k_x x+k_y y)),
\end{equation}
where $k_x $ and $k_y $ are real constants, $\mbox{i}$ is the imaginary 
unit, and the boundary conditions are periodic. Suppose that the exact 
solution of this PDE has the form of
\begin{equation}
\label{eq22}
f(x,y,t)=\exp (\omega t)f(x,y,0),
\end{equation}
where $\omega $ is a complex constant. Substituting this into Eq.~(\ref{eq20}) 
yields
\begin{equation}
\label{eq23}
\omega =\mbox{i}^{m+n}\mu k_x ^mk_y ^n.
\end{equation}

The characteristic of this exact solution can roughly be categorized into 
the following two solutions:

\noindent
\quad For $m+n=2M$:
\begin{equation}
\label{eq24}
f=\exp ((-1)^M\mu k_x ^mk_y ^nt)\exp (\mbox{i}(k_x x+k_y y)),
\end{equation}

\noindent
\quad For $m+n=2M+1$:
\begin{equation}
\label{eq25}
f=\exp (\mbox{i}(-1)^M\mu k_x ^mk_y ^nt)\exp (\mbox{i}(k_x x+k_y y)),
\end{equation}
where $M=0,1,2,\cdots $. The former represents exponential decay or growth 
of the solution, whereas the latter represents phase shift without changing 
amplitude.

Equation (\ref{eq24}) suggests that in order for a stable finite difference scheme 
for $m+n=2M$ to exist, the {\it amplification exponent} of solution (\ref{eq24}),
\begin{equation}
\label{eq26}
\alpha (m,n;\mu )=(-1)^M\mu k_x ^mk_y ^n,
\end{equation}
must be zero or less for any value of wavenumbers. Furthermore, if $m$ and 
$n$ are even numbers, $k_x ^mk_y ^n$ in this exponent is always positive and 
the stability is thus determined by the sign of $(-1)^M\mu $. (If, for 
example, $m=2$ and $n=0$, which results in $M=1$, the PDEs with $\mu \ge 0$ 
are numerically stable, while those with $\mu <0$ are unconditionally 
unstable, a conclusion that is consistent with the well-known fact that a 
positive diffusion equation can be solved stably but a negative one can 
not.) Otherwise, i.e., if $m$ and $n$ are odd numbers, $k_x ^mk_y ^n$ can be 
either positive or negative, meaning that the PDEs in this case are always 
unconditionally unstable because modes in any direction can appear in 
turbulent flows. (This conclusion is consistent with the known fact that the 
PDE for $m=n=1$ is unconditionally unstable; see, e.g., Refs.~\cite{ref23,ref17,ref18}). 
For $(-1)^M\mu <0$, for example, the modes of $k_x k_y <0$ must be unstable, 
implying that if the sign of $\mu $ is constant, the PDEs in this case 
always exhibit numerical instability in a fixed range of directions. This 
finding plays an important role in our main subject discussed in the next 
section.

On the other hand, Eq.~(\ref{eq25}) suggests that if $m+n$ is an odd number, a 
stable finite difference scheme should always exist because the 
amplification exponent
\begin{equation}
\label{eq27}
\alpha (m,n;\mu )=\mbox{i}(-1)^M\mu k_x ^mk_y ^n
\end{equation}
has an imaginary value and hence the absolute value of the amplification 
factor, $\vert \exp (\mbox{i}(-1)^M\mu k_x ^mk_y ^nt)\vert $, is unity. 
Indeed, the advection equations (corresponding to the case of, e.g., $m=1$ 
with $n=0)$ and the Korteweg-de Vries (KdV) equations involving third- 
and/or fifth-order dispersion terms (e.g., for $m=1,3,5$ with $n=0)$ have 
been solved stably and accurately by finite difference schemes; see, e.g., 
Refs.~\cite{ref32,ref33,ref34} for recent progress in finite difference schemes for KdV 
equations.

The present theoretical results can be summarized as follows: A stable 
finite difference solver must exist for odd-order PDEs (i.e., when $m+n$ is 
an odd number). For even-order PDEs, a stable solver exists only if both $m$ 
and $n$ are even numbers and $(-1)^{(m+n)/2}\mu $ is negative; otherwise, 
the PDE is unconditionally unstable by any finite difference scheme, since 
the numerical perturbations grow exponentially in numerical simulations. 
This conclusion may also be true for the finite volume, finite element, and 
compact difference schemes.

The total amplification exponent $\alpha _T $ of a complicated PDE,
\begin{equation}
\label{eq28}
\frac{\partial f}{\partial t}=\sum\limits_i {\mu _i \left( {\frac{\partial 
}{\partial x}} \right)^{m_i }\left( {\frac{\partial }{\partial y}} 
\right)^{n_i }f} ,
\end{equation}
can be determined by
\begin{equation}
\label{eq29}
\alpha _T =\sum\limits_i {{\rm Re}[\alpha (m_i ,n_i ;\mu _i )]} .
\end{equation}
Though this is, for variable $\mu _i $, only an approximation, it should 
work sufficiently in many situations.

\section{Stability analysis of the filtered system}
\label{secIV}
We present several analytical results for the numerical stability of a 
filtered system determined by combining the stability analysis described in 
the previous section with the exact expansion series (\ref{eq13}) and (\ref{eq14}) or the 
identity given using integration by parts, (\ref{eq15}). As stated in Sec.~\ref{secII}, we 
consider up to 2D cases for the sake of simplicity, and only analyze the 
stability of Eq.~(\ref{eq12}).

\subsection{1D mean velocity cases}
Suppose that
\[
U_1 =U_1 (x_2 ) \quad \mbox{and} \quad U_2 =U_3 =0,
\]
(this means that the velocity satisfies the divergence-free condition), 
which leads to
\begin{equation}
\label{eq30}
h_1 =U_1 \frac{\partial u'_1 }{\partial x_1 }+\frac{\partial U_1 
}{\partial x_2 }u'_2 ,
\end{equation}
\begin{equation}
\label{eq31}
h_2 =U_1 \frac{\partial u'_2 }{\partial x_1 },
\end{equation}
\begin{equation}
\label{eq32}
h_3 =U_1 \frac{\partial u'_3 }{\partial x_1 }.
\end{equation}
Because $U_1 $ depends only on $x_2 $, filtering in the $x_1 $ and $x_3 $ 
directions (i.e., in the homogeneous directions) results in
\begin{equation}
\label{eq33}
{\mathcal R}_{13} [h_i ]=0\quad \mbox{for }i=1,2,3,
\end{equation}

When the Gaussian filter in the $x_2 $ direction is applied to 
Eqs.~(\ref{eq30})-(\ref{eq32}), some mathematical manipulations are needed to obtain a closed 
formula because $U_1 $ and $\partial U_1 /\partial x_2 $ cannot simply be 
put outside the convolution operation. We consider here the case where $U_1 
$ is described by a finite-order polynomial:
\begin{equation}
\label{eq34}
U_1 (x_2 )=\sum\limits_{n=0}^N {a_n x_2 ^n} ,
\end{equation}
where $a_n $ ($n=0,1,\ldots ,N)$ are real constants and $N$ is the order of 
this polynomial. Using the 1D expansion series (\ref{eq13}), we have
\begin{equation}
\label{eq35}
{\mathcal R}_2 [h_1 ]={\mathcal L}_{[1]} \bar u'_1 +{\mathcal L}_{[2]} \bar u'_2 ,
\end{equation}
\begin{equation}
\label{eq36}
{\mathcal R}_2 [h_2 ]={\mathcal L}_{[1]} \bar u'_2 ,
\end{equation}
\begin{equation}
\label{eq37}
{\mathcal R}_2 [h_3 ]={\mathcal L}_{[1]} \bar u'_3 ,
\end{equation}
where
\begin{equation}
\label{eq38}
{\mathcal L}_{[1]} \equiv \sum\limits_{n=1}^\infty {\frac{1}{n!}\left( {\frac{\Delta _2 
^2}{2\gamma }} \right)^n\frac{\partial ^n\bar {U}_1 }{\partial x_2 
^n}\frac{\partial ^{n+1}}{\partial x_1 \partial x_2 ^n}} ,
\end{equation}
\begin{equation}
\label{eq39}
{\mathcal L}_{[2]} \equiv \sum\limits_{n=1}^\infty {\frac{1}{n!}\left( {\frac{\Delta _2 
^2}{2\gamma }} \right)^n\frac{\partial ^{n+1}\bar {U}_1 }{\partial x_2 
^{n+1}}\frac{\partial ^n}{\partial x_2 ^n}} .
\end{equation}
In the following, we examine some special cases for deriving the stability 
conditions for the 1D flows.

If $N=1$, operators (\ref{eq38}) and (\ref{eq39}), respectively, reduce to
\begin{equation}
\label{eq40}
{\mathcal L}_{[1]} =\left( {\frac{\Delta _2 ^2}{2\gamma }} \right)\frac{\partial \bar 
{U}_1 }{\partial x_2 }\frac{\partial ^2}{\partial x_1 \partial x_2 },
\end{equation}
\begin{equation}
\label{eq41}
{\mathcal L}_{[2]} =0,
\end{equation}
and Eq.~(\ref{eq12}) then becomes
\begin{equation}
\label{eq42}
\left( {\nu \frac{\partial ^2}{\partial x_1 ^2}+\nu \frac{\partial 
^2}{\partial x_2 ^2}-{\mathcal L}_{[1]} } \right)\bar u'_i ,\quad i=1,2,3.
\end{equation}
Here the diffusion operator in the $x_3 $ direction is neglected because it 
does not alter the resulting stability condition that is applicable to any 
wavenumber; note that the neglected operator does not stabilize the modes in 
the $(x_1 ,x_2 )$ plane but that ${\mathcal L}_{[1]} $, being unstable, only has 
derivatives with respect to $x_1 $ and $x_2 $. Based on Eqs.~(\ref{eq26}) and (\ref{eq29}), 
we have
\begin{equation}
\label{eq43}
\alpha _T \simeq -\nu (k_x ^2+k_y ^2)+\left( {\frac{\Delta _2 ^2}{2\gamma }} 
\right)\frac{\partial \bar {U}_1 }{\partial x_2 }k_x k_y .
\end{equation}
Substituting $(k_x ,k_y )=\left| {\rm {\bf k}} \right|(\cos \theta ,\sin 
\theta )$ with $\theta \in [-\pi ,\pi ]$ into $\alpha _T \le 0$ yields
\begin{equation}
\label{eq44}
1\ge \max \left( {\frac{\Delta _2 ^2}{4\gamma \nu }\frac{\partial \bar {U}_1 
}{\partial x_2 }\sin 2\theta } \right)=\frac{\Delta _2 ^2}{4\gamma \nu 
}\left| {\frac{\partial \bar {U}_1 }{\partial x_2 }} \right|.
\end{equation}
This stability condition is equivalent to that for linear shears determined 
in Ref.~\cite{ref19} by a different approach, which imposed a strong restriction on 
the choice of the wall-normal filter width for use in the viscous sublayer 
in plane channel flows.

For $N=2$, ${\mathcal L}_{[1]} $ and ${\mathcal L}_{[2]} $ become
\begin{equation}
\label{eq45}
{\mathcal L}_{[1]} =\left( {\frac{\Delta _2 ^2}{2\gamma }} \right)\frac{\partial \bar 
{U}_1 }{\partial x_2 }\frac{\partial ^2}{\partial x_1 \partial x_2 
}+\frac{1}{2}\left( {\frac{\Delta _2 ^2}{2\gamma }} \right)^2\frac{\partial 
^2\bar {U}_1 }{\partial x_2 ^2}\frac{\partial ^3}{\partial x_1 \partial x_2 
^2},
\end{equation}
\begin{equation}
\label{eq46}
{\mathcal L}_{[2]} =\left( {\frac{\Delta _2 ^2}{2\gamma }} \right)\frac{\partial ^2\bar 
{U}_1 }{\partial x_2 ^2}\frac{\partial }{\partial x_2 }.
\end{equation}
As proven in Sec.~\ref{secIII}, the third-order operator in Eq.~(\ref{eq45}) can be ignored 
in stability analysis. Moreover, ${\mathcal L}_{[2]} \bar u'_2 $ in Eq.~(\ref{eq35}) can 
also be neglected, because the numerical stability of $\bar u'_2 $ is 
determined independently of Eq.~(\ref{eq35}), by Eq.~(\ref{eq36}), and 
furthermore, if Eq.~(\ref{eq36}) is unstable, then Eq.~(\ref{eq35}) should 
be unstable. Therefore, the stability 
condition in the present example is the same as that in the previous case, 
Eq.~(\ref{eq44}).

For $N=3$, the total amplification exponent for $\nu (\partial ^2/\partial 
x_1 ^2+\partial ^2 /\partial x_2 ^2)-{\mathcal L}_{[1]} $ is
\begin{equation}
\label{eq47}
\alpha _T \simeq -\nu (k_x ^2+k_y ^2)+\left( {\frac{\Delta _2 ^2}{2\gamma }} 
\right)\frac{\partial \bar {U}_1 }{\partial x_2 }k_x k_y -\frac{1}{6}\left( 
{\frac{\Delta _2 ^2}{2\gamma }} \right)^3\frac{\partial ^3\bar {U}_1 
}{\partial x_2 ^3}k_x k_y ^3.
\end{equation}
Using this and $(k_x ,k_y )=\left| {\rm {\bf k}} \right|(\cos \theta ,\sin 
\theta )$, we obtain the following stability condition:
\begin{equation}
\label{eq48}
1\ge \frac{1}{2\nu }\left( {\frac{\Delta _2 ^2}{2\gamma }} 
\right)\frac{\partial \bar {U}_1 }{\partial x_2 }s-\frac{1}{24\nu }\left( 
{\frac{\Delta _2 ^2}{2\gamma }} \right)^3\frac{\partial ^3\bar {U}_1 
}{\partial x_2 ^3}\left| {\rm {\bf k}} \right|^2s(1\pm \sqrt {1-s^2} ),
\end{equation}
\[
s\equiv \sin 2\theta \in [-1,1].
\]
This has to be fulfilled for any choice of $\left| {\rm {\bf k}} \right|$ 
and $\theta $.

To show how the restriction (\ref{eq48}) works in a realistic situation, we consider 
here the inertial sublayer forming in a plane channel flow. Following Dean 
\cite{ref35}, the streamwise mean velocity in the inertial sublayer is approximately 
described by
\begin{equation}
\label{eq49}
\frac{U_1 (x_2 ^+)}{u_\tau }=2.44\ln (x_2 ^+)+5.17,
\end{equation}
where $u_\tau $ is the wall-friction velocity and $x_2 ^+=x_2 u_\tau /\nu $ 
is the distance from the plane wall in wall units. For $30\le x_2^+ \le 80$, 
Eq.~(\ref{eq49}) can be well approximated by a cubic polynomial:
\begin{equation}
\label{eq50}
\frac{U_1 (x_2^+ )}{u_\tau }=\sum\limits_{n=0}^3 {a_n x_2^{+\,n} } ,
\end{equation}
\[
\left\{ {{\begin{array}{*{20}l}
 {a_0 =10.5,} & {a_1 =0.134,} \\
 {a_2 =-1.23\times 10^{-3},} & {a_3 =5\times 10^{-6}.} \\
\end{array} }} \right.
\]
Using this and Eqs.~(\ref{eq16}) and (\ref{eq17}), the filtered derivatives in Eq.~(\ref{eq48}) are 
determined as
\begin{equation}
\label{eq51}
\frac{\partial \bar {U}_1 }{\partial x_2 }=u_\tau \frac{u_\tau }{\nu }\left[ 
{a_1 +2a_2 x_2 ^++3a_3 \left( {x_2 ^{+\,2}+\frac{\Delta _2 ^{+\,2}}{2\gamma 
}} \right)} \right],
\end{equation}
\begin{equation}
\label{eq52}
\frac{\partial ^3\bar {U}_1 }{\partial x_2 ^3}=u_\tau \left( {\frac{u_\tau 
}{\nu }} \right)^36a_3 ,
\end{equation}
where $\Delta _2 ^+=\Delta _2 u_\tau /\nu $. Substituting them and $\gamma 
=6$ into Eq.~(\ref{eq48}) yields
\begin{equation}
\label{eq53}
1\ge \left( {\frac{A_{[1]} \Delta _2 ^{+\,2}}{24}+\frac{a_3 \Delta _2 
^{+\,4}}{96}} \right)s-\frac{a_3 \Delta _2 ^{+\,6}}{6912}\left| {\rm {\bf 
k}} \right|^2\left( {\frac{u_\tau }{\nu }} \right)^{-2}s(1\pm \sqrt {1-s^2} 
),
\end{equation}
where
\[
A_{[1]} =a_1 +2a_2 x_2 ^++3a_3 x_2 ^{+\,2}.
\]
Assuming that $\max (\left| {\rm {\bf k}} \right|)\Delta _2 =\pi $, i.e., 
the maximum resolved wavenumber is determined by the Nyquist wavenumber 
based on the wall-normal filter width, Eq.~(\ref{eq53}) can be further rewritten as
\begin{equation}
\label{eq54}
1\ge \left( {\frac{A_{[1]} \Delta _2 ^{+\,2}}{24}+\frac{a_3 \Delta _2 
^{+\,4}}{96}} \right)s-\frac{a_3 \pi ^2\Delta _2 ^{+\,4}}{6912}s(1\pm \sqrt 
{1-s^2} ).
\end{equation}
At $x_2 ^+=55$, for instance, this becomes
\begin{equation}
\label{eq55}
1\ge (1.85\times 10^{-3}\Delta ^{+\,2}+5.21\times 10^{-8}\Delta 
^{+\,4})s-7.14\times 10^{-9}\Delta ^{+\,4}s(1\pm \sqrt {1-s^2} ).
\end{equation}
From this, for $\Delta ^+=10$, $20$, and $25$, respectively, we have
\begin{equation}
\label{eq56}
1\ge 0.185s-7.14\times 10^{-5}s(1\pm \sqrt {1-s^2} ),
\end{equation}
\begin{equation}
\label{eq57}
1\ge 0.748s-0.00114s(1\pm \sqrt {1-s^2} ),
\end{equation}
\begin{equation}
\label{eq58}
1\ge 1.18s-0.00279s(1\pm \sqrt {1-s^2} ).
\end{equation}
The first two are true for any $s\in [-1,1]$, but the last is not. (Note 
that $\max _s [s(1\pm \sqrt {1-s^2} )]=-\min _s [s(1\pm \sqrt {1-s^2} 
)]=3\sqrt 3 /4\simeq 1.30$.) The filter width suggested here for stability 
is comparable to that used in actual channel flow computations.

Based on the stability analysis described in Sec.~\ref{secIII}, it is found that for 
larger $N$, all of the even-order differential operators in Eq.~(\ref{eq38}) are 
unstable, whereas the odd-order ones have no influence on the numerical 
stability. That is, the high-order terms do not help stability. This result 
suggests that in most cases of 1D shear, the subfilter-scale stress terms 
would be unstable, thus leading to a divergence of numerical solution, if an 
unsuitably large filter width is used.

\subsection{2D mean velocity cases}
Next, we consider 2D problems. Suppose that
\[
U_1 =U_1 (x_1 ,x_2 ), \quad U_2 =U_2 (x_1 ,x_2 ), \quad \mbox{and} \quad U_3 =0,
\]
resulting in
\begin{equation}
\label{eq59}
h_1 ={\mathcal D}u'_1 +\frac{\partial U_1 }{\partial x_1 }u'_1 +\frac{\partial U_1 
}{\partial x_2 }u'_2 ,
\end{equation}
\begin{equation}
\label{eq60}
h_2 ={\mathcal D}u'_2 +\frac{\partial U_2 }{\partial x_1 }u'_1 +\frac{\partial U_2 
}{\partial x_2 }u'_2 ,
\end{equation}
\begin{equation}
\label{eq61}
h_3 ={\mathcal D}u'_3 ,
\end{equation}
and
\begin{equation}
\label{eq62}
\frac{\partial U_i }{\partial x_i }=\frac{\partial U_1 }{\partial x_1 
}+\frac{\partial U_2 }{\partial x_2 }=0.
\end{equation}
Here we introduced an advection operator,
\[
{\mathcal D}=U_1 \frac{\partial }{\partial x_1 }+U_2 \frac{\partial }{\partial x_2 }.
\]
Because the mean velocity is independent of $x_3 $, we know that
\begin{equation}
\label{eq63}
{\mathcal R}_3 [h_i ]=0\quad \mbox{for }i=1,2,3,
\end{equation}
and consequently the resulting formulas of the residual stresses for 
${\mathcal G}_{123} \star $ and for ${\mathcal G}_{12} \star $ are the same, allowing for the 
consideration based on 2D filtering. However, even in 2D, the complete set 
of the closed residual forces derived using Eq.~(\ref{eq14}) is intricate and 
inconvenient for theoretical analysis, and hence we only consider some 
simple cases.

The first example assumes that the mean velocities are described locally by
\begin{equation}
\label{eq64}
U_1 =bx_1 \quad \mbox{and}\quad U_2 =-bx_2 ,
\end{equation}
where $b$ is a positive constant, that is, stretches uniformly in the $x_1 $ 
direction. Here, the term ``locally'' means ``in a region sufficiently 
larger than the filter widths.'' These assumptions reduce Eqs.~(\ref{eq59})-(\ref{eq61}) to
\begin{equation}
\label{eq65}
h_1 ={\mathcal D}u'_1 +bu'_1 ,
\end{equation}
\begin{equation}
\label{eq66}
h_2 ={\mathcal D}u'_2 -bu'_2 ,
\end{equation}
\begin{equation}
\label{eq67}
h_3 ={\mathcal D}u'_3 .
\end{equation}
Since, in this case,
\[
{\mathcal G}_j \star \left( {U_k \frac{\partial u'_i }{\partial x_k }} \right)=U_k 
\frac{\partial ({\mathcal G}_j \star u'_i )}{\partial x_k }\quad \mbox{for }j\ne 
k,\quad \mbox{no summation over }k
\]
and
\[
{\mathcal R}_{12} [bu'_1 ]={\mathcal R}_{12} [bu'_2 ]=0
\]
are true, the residual forces are expressed as
\begin{equation}
\label{eq68}
{\mathcal R}_{12} [h_i ]={\mathcal R}_1 \left[ {bx_1 \frac{\partial ({\mathcal G}_2 \star u'_i )}{\partial 
x_1 }} \right]-{\mathcal R}_2 \left[ {bx_2 \frac{\partial ({\mathcal G}_1 \star u'_i )}{\partial 
x_2 }} \right],\quad i=1,2,3.
\end{equation}
Using the 1D expansion series (\ref{eq13}) or integration by parts (\ref{eq15}), Eq.~(\ref{eq68}) is 
rewritten into the closed form,
\begin{equation}
\label{eq69}
{\mathcal R}_{12} [h_i ]=b\frac{\Delta _1 ^2}{2\gamma }\frac{\partial ^2\bar u'_i 
}{\partial x_1 ^2}-b\frac{\Delta _2 ^2}{2\gamma }\frac{\partial ^2\bar 
u'_i }{\partial x_2 ^2}.
\end{equation}
This acts as negative diffusion in the $x_1 $ direction (i.e., in the 
stretching direction) but as positive diffusion in the $x_2 $ direction, a 
result that is consistent with Leonard's finding shown in the studies on 
tensor-diffusivity models \cite{ref16,ref13}. The amplification exponent of the 
difference between the viscosity term and ${\mathcal R}_{12} [h_i ]$ is
\begin{equation}
\label{eq70}
\alpha _T \simeq -\left( {\nu -b\frac{\Delta _1 ^2}{2\gamma }} \right)k_x 
^2-\left( {\nu +b\frac{\Delta _2 ^2}{2\gamma }} \right)k_y ^2,
\end{equation}
which expression leads to the stability condition
\begin{equation}
\label{eq71}
\Delta _1 \le \frac{\sqrt {2\gamma \nu } }{\sqrt b }.
\end{equation}
Equation (\ref{eq71}) indicates that a smaller filter width is needed for stronger 
stretching, and the largest filter width usable in the stretching direction 
is inversely proportional to the square root of the velocity gradient.

\begin{figure}[htbp]
\includegraphics[width=8cm]{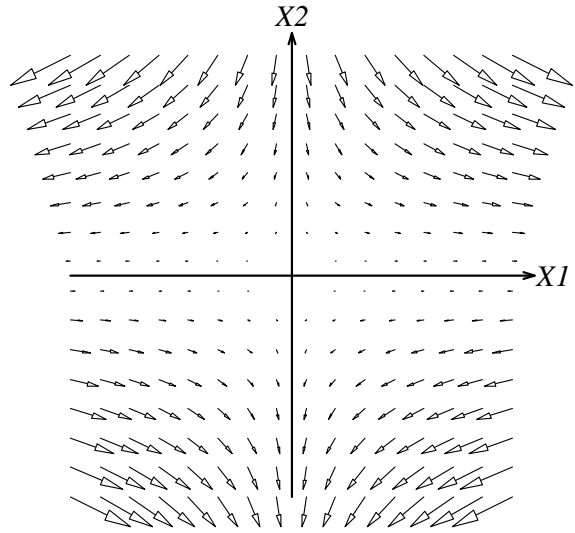}
\caption
{Vector plot of the flow field described by Eqs.~(\ref{eq72}) and (\ref{eq73}) in 
arbitrary units; note that this flow field is self-similar with respect to 
constant multiplication of the coordinates, $(x_1 ,x_2 )\to (cx_1 ,cx_2 )$, 
where $c$ is a real constant. The center point of the coordinate system 
shows the origin $(x_1 ,x_2 )=(0,0)$.}
\label{fig1}
\end{figure}

The next example is complicated; not only the normal stresses but also a 
shear stress appear in the mean field. Suppose that $U_1 $ is described 
locally by
\begin{equation}
\label{eq72}
U_1 (x_1 ,x_2 )=bx_1 x_2 ,
\end{equation}
which involves both normal and shear stresses. Then one has, from the 
divergence-free condition,
\begin{equation}
\label{eq73}
U_2 (x_2 )=-\frac{b}{2}x_2 ^2,
\end{equation}
where $U_2 (0)=0$ is assumed without loss of generality, and $b$ denotes a 
positive constant as in the previous example; see Fig.~\ref{fig1} showing the vector 
plot of this velocity field around the origin $(x_1 ,x_2 )=(0,0)$. The upper 
side ($x_2 \ge 0)$ of this figure seems to represent a flow impinging on the 
wall located at $x_2 =0$. Substituting Eqs.~(\ref{eq72}) and (\ref{eq73}), Eqs.~(\ref{eq59})-(\ref{eq61}) 
become
\begin{equation}
\label{eq74}
h_1 ={\mathcal D}u'_1 +bx_2 u'_1 +bx_1 u'_2 ,
\end{equation}
\begin{equation}
\label{eq75}
h_2 ={\mathcal D}u'_2 -bx_2 u'_2 ,
\end{equation}
\begin{equation}
\label{eq76}
h_3 ={\mathcal D}u'_3 .
\end{equation}
Since
\[
{\mathcal G}_{12} \star \left( {U_1 \frac{\partial u'_i }{\partial x_1 }} \right)=b{\mathcal G}_1 
\star \left[ {x_1 {\mathcal G}_2 \star \left( {x_2 \frac{\partial u'_i }{\partial x_1 
}} \right)} \right],
\]
Eq.~(\ref{eq15}) (and also (\ref{eq13})) can be used to obtain
\begin{equation}
\label{eq77}
{\mathcal R}_{12} \left[ {U_1 \frac{\partial u'_i }{\partial x_1 }} 
\right]=\frac{\Delta _1 ^2}{2\gamma }bx_2 \frac{\partial ^2\bar u'_i 
}{\partial x_1 ^2}+\frac{\Delta _2 ^2}{2\gamma }bx_1 \frac{\partial ^2\bar 
u'_i }{\partial x_1 \partial x_2 }+\frac{\Delta _2 ^2}{2\gamma 
}\frac{\Delta _1 ^2}{2\gamma }b\frac{\partial ^3\bar u'_i }{\partial x_1 
^2\partial x_2 },
\end{equation}
where we used ${\mathcal G}_{12} \star U_1 =U_1 $. Furthermore, using Eq.~(\ref{eq13}) yields
\begin{eqnarray}
\label{eq78}
{\mathcal R}_{12} \left[ {U_2 \frac{\partial u'_i }{\partial x_2 }} \right] &=& {\mathcal R}_2 
\left[ {U_2 \frac{\partial ({\mathcal G}_1 \star u'_i )}{\partial x_2 }} \right] \nonumber \\ 
 &=& -\frac{\Delta _2 ^2}{2\gamma }bx_2 \frac{\partial ^2\bar u'_i 
}{\partial x_2 ^2}-\frac{1}{2}\left( {\frac{\Delta _2 ^2}{2\gamma }} 
\right)^2b\frac{\partial ^3\bar u'_i }{\partial x_2 ^3},
\end{eqnarray}
where we used
\begin{equation}
\label{eq79}
\frac{\partial \bar {U}_2 }{\partial x_2 }=\frac{\partial U_2 }{\partial x_2 
}=-bx_2 \quad \mbox{and}\quad \frac{\partial ^2\bar {U}_2 }{\partial x_2 
^2}=\frac{\partial ^2U_2 }{\partial x_2 ^2}=-b.
\end{equation}
The remaining terms can also be rewritten into a closed form using Eq.~(\ref{eq15}) 
or (\ref{eq13}). Finally, we obtain the closed residual stresses:
\begin{equation}
\label{eq80}
{\mathcal R}_{12} [h_1 ]={\mathcal L}_{[3]} \bar u'_1 +b\frac{\Delta _2 ^2}{2\gamma 
}\frac{\partial \bar u'_1 }{\partial x_2 }+b\frac{\Delta _1 ^2}{2\gamma 
}\frac{\partial \bar u'_2 }{\partial x_1 },
\end{equation}
\begin{equation}
\label{eq81}
{\mathcal R}_{12} [h_2 ]={\mathcal L}_{[3]} \bar u'_2 -b\frac{\Delta _2 ^2}{2\gamma 
}\frac{\partial \bar u'_2 }{\partial x_2 },
\end{equation}
\begin{equation}
\label{eq82}
{\mathcal R}_{12} [h_3 ]={\mathcal L}_{[3]} \bar u'_3 ,
\end{equation}
where
\begin{eqnarray}
\label{eq83}
{\mathcal L}_{[3]} &\equiv& bx_2 \left( {\frac{\Delta _1 ^2}{2\gamma }\frac{\partial 
^2}{\partial x_1 ^2}-\frac{\Delta _2 ^2}{2\gamma }\frac{\partial 
^2}{\partial x_2 ^2}} \right)+\frac{\Delta _2 ^2}{2\gamma }bx_1 
\frac{\partial ^2}{\partial x_1 \partial x_2 } \nonumber \\ 
 &+& \frac{\Delta _2 ^2}{2\gamma }b\left( {\frac{\Delta _1 ^2}{2\gamma 
}\frac{\partial ^3}{\partial x_1 ^2\partial x_2 }-\frac{1}{2}\frac{\Delta _2 
^2}{2\gamma }\frac{\partial ^3}{\partial x_2 ^3}} \right).
\end{eqnarray}
In ${\mathcal L}_{[3]} $ we can see various kinds of operators: negative and positive 
diffusions, second- and third-order cross derivatives, and third-order 
dispersion, among which third-order terms do not alter the amplification 
exponent $\alpha _T $. Also, the last two terms of Eq.~(\ref{eq80}) and the last of 
Eq.~(\ref{eq81}), first-order derivatives, may not concern the numerical stability. 
That is, the differential operators responsible for stability are
\begin{equation}
\label{eq84}
\nu \left( {\frac{\partial ^2}{\partial x_1 ^2}+\frac{\partial ^2}{\partial 
x_2 ^2}} \right)-bx_2 \left( {\frac{\Delta _1 ^2}{2\gamma }\frac{\partial 
^2}{\partial x_1 ^2}-\frac{\Delta _2 ^2}{2\gamma }\frac{\partial 
^2}{\partial x_2 ^2}} \right)-\frac{\Delta _2 ^2}{2\gamma }bx_1 
\frac{\partial ^2}{\partial x_1 \partial x_2 },
\end{equation}
whose amplification exponent is
\begin{equation}
\label{eq85}
\alpha _T \simeq -\nu (k_x ^2+k_y ^2)+bx_2 \left( {\frac{\Delta _1 
^2}{2\gamma }k_x ^2-\frac{\Delta _2 ^2}{2\gamma }k_y ^2} 
\right)+\frac{\Delta _2 ^2}{2\gamma }bx_1 k_x k_y .
\end{equation}
From this, the stability condition of the present example is determined as
\begin{equation}
\label{eq86}
1\ge \frac{b}{4\gamma \nu }[(\Delta _1 ^2-\Delta _2 ^2)x_2 +(\Delta _1 
^2+\Delta _2 ^2)x_2 \cos 2\theta +\Delta _2 ^2x_1 \sin 2\theta ],
\end{equation}
which should be fulfilled for any choice of $\theta $.

Let us consider some particular cases to show how Eq.~(\ref{eq86}) works. The 2D 
formula can be used to discover the stability conditions for 1D filtering as 
well, because $\mathop {\lim }\limits_{\Delta _i \to 0} ({\mathcal G}_i \star f)=f$. For 
$\Delta _2 \to 0$ and $\Delta _1 \to 0$, respectively, Eq.~(\ref{eq86}) reduces to
\begin{equation}
\label{eq87}
1\ge \frac{b}{4\gamma \nu }\left( {x_2 +x_2 \cos 2\theta } \right)\Delta _1 
^2,
\end{equation}
\begin{equation}
\label{eq88}
1\ge \frac{b}{4\gamma \nu }\left( {-x_2 +x_2 \cos 2\theta +x_1 \sin 2\theta 
} \right)\Delta _2 ^2.
\end{equation}
On the other hand, if the condition $\Delta _1 =\Delta _2 =\Delta $ needs to 
be satisfied for some factor, then Eq.~(\ref{eq86}) becomes
\begin{equation}
\label{eq89}
1\ge \frac{b}{4\gamma \nu }\left( {2x_2 \cos 2\theta +x_1 \sin 2\theta } 
\right)\Delta ^2.
\end{equation}
Below we briefly discuss these three cases.

For $x_1 =x_2 >0$, Eqs.~(\ref{eq87})-(\ref{eq89}) reduce to
\begin{equation}
\label{eq90}
1\ge \max \left[ {\frac{bx_2 }{4\gamma \nu }\left( {1+\cos 2\theta } 
\right)\Delta _1 ^2} \right]=\left( {\frac{bx_2 }{4\gamma \nu }} 
\right)2\Delta _1 ^2,
\end{equation}
\begin{equation}
\label{eq91}
1\ge \max \left[ {\frac{bx_2 }{4\gamma \nu }(-1+\cos 2\theta +\sin 2\theta 
)\Delta _2 ^2} \right]=\left( {\frac{bx_2 }{4\gamma \nu }} \right)(\sqrt 2 
-1)\Delta _2 ^2,
\end{equation}
and
\begin{equation}
\label{eq92}
1\ge \max \left[ {\frac{bx_2 }{4\gamma \nu }\left( {2\cos 2\theta +\sin 
2\theta } \right)\Delta ^2} \right]=\left( {\frac{bx_2 }{4\gamma \nu }} 
\right)\sqrt 5 \Delta ^2,
\end{equation}
and the respective influential differential operators are
\begin{equation}
\label{eq93}
\nu \left( {\frac{\partial ^2}{\partial x_1 ^2}+\frac{\partial ^2}{\partial 
x_2 ^2}} \right)-bx_2 \frac{\Delta _1 ^2}{2\gamma }\frac{\partial 
^2}{\partial x_1 ^2},
\end{equation}
\begin{equation}
\label{eq94}
\nu \left( {\frac{\partial ^2}{\partial x_1 ^2}+\frac{\partial ^2}{\partial 
x_2 ^2}} \right)+bx_2 \frac{\Delta _2 ^2}{2\gamma }\frac{\partial 
^2}{\partial x_2 ^2}-bx_2 \frac{\Delta _2 ^2}{2\gamma }\frac{\partial 
^2}{\partial x_1 \partial x_2 },
\end{equation}
and
\begin{equation}
\label{eq95}
\nu \left( {\frac{\partial ^2}{\partial x_1 ^2}+\frac{\partial ^2}{\partial 
x_2 ^2}} \right)-bx_2 \left( {\frac{\Delta ^2}{2\gamma }\frac{\partial 
^2}{\partial x_1 ^2}-\frac{\Delta ^2}{2\gamma }\frac{\partial ^2}{\partial 
x_2 ^2}} \right)-bx_2 \frac{\Delta ^2}{2\gamma }\frac{\partial ^2}{\partial 
x_1 \partial x_2 }.
\end{equation}
Among the stability conditions, Eq.~(\ref{eq91}) gives the weakest restriction on 
the filter width; the second-to-last term of Eq.~(\ref{eq94}), being a positive 
diffusion term resulting from compression in the $x_2 $ direction, mitigates 
the instability of the last term, the cross derivative resulting from the 
shear in the same direction. In contrast, Eq.~(\ref{eq92}) is the most restrictive 
condition, resulting from the coexistence of a negative-diffusion term and a 
cross derivative term.

For $x_1 =x_2 <0$, on the other hand, Eqs.~(\ref{eq87})-(\ref{eq89}) become
\begin{equation}
\label{eq96}
1\ge 0\times \Delta _1 ^2,
\end{equation}
\begin{equation}
\label{eq97}
1\ge \left( {\frac{-bx_2 }{4\gamma \nu }} \right)(1+\sqrt 2 )\Delta _2 ^2,
\end{equation}
\begin{equation}
\label{eq98}
1\ge \left( {\frac{-bx_2 }{4\gamma \nu }} \right)\sqrt 5 \Delta ^2.
\end{equation}
In this case, Eqs.~(\ref{eq97}) and (\ref{eq98}), whose respective influential derivatives 
have both negative-diffusion and cross-derivative terms, give restrictions 
of almost equal strength, while Eq.~(\ref{eq96}), involving positive-diffusion terms 
only, imposes no restriction. The results provided here denote that the 
numerical stability of the filtered system and the possible choice of filter 
widths depend on how the filtering operations are applied; this conclusion 
confirms the same assertion presented in Ref.~\cite{ref19}.

\section{Discussion of the negative-diffusion term}
\label{secV}
As has been shown in the previous sections, many kinds of numerically 
unstable terms are derived by the Gaussian filtering operation. In this 
section we would like to remark on the negative-diffusion term appearing in 
the stretching direction to clarify why such unstable terms appear and how 
the terms work. The discussion also clarifies that the present suggestions 
are basically true even for other smooth filters.

\begin{figure}[htbp]
\includegraphics[width=8cm]{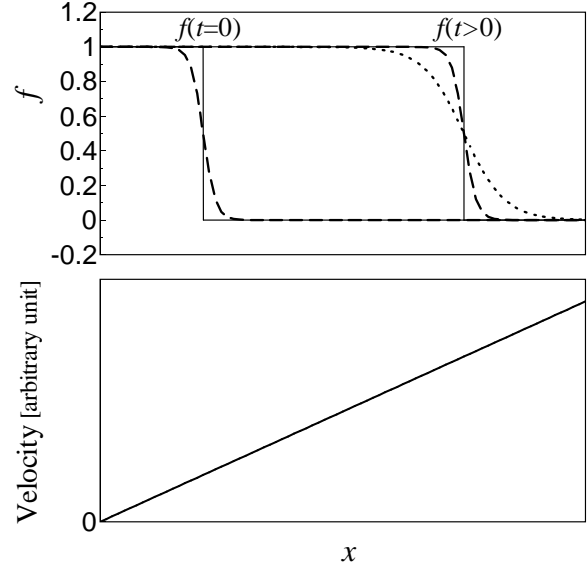}
\caption
{Physical meaning of the negative diffusivity in the pure advection 
of a discontinuous step function in a stretching field. The solid lines 
in the upper figure 
denote $f(t=0)$ and $f(t=t_1 >0)$, and the dashed curves denote $\bar 
{f}(t=0)$ and $\bar {f}(t=t_1 >0)$. If $\bar {f}$ is advanced using a pure 
advection equation, its characteristic width gradually expands, as shown by 
the dots. A negative diffusion term must appear in the filtered advection 
equation to counteract this artificial expansion.}
\label{fig2}
\end{figure}

In Ref.~\cite{ref13}, Leonard considered the pure advection of a sinusoidal wave in 
a stretching velocity field where the amplitude of the (unfiltered) 
sinusoidal wave remains constant, and provided an interpretation of the 
negative diffusivity of the tensor-diffusivity model as follows: As a 
sinusoidal wave propagates into a stretching velocity field, its wavenumber 
gradually decreases, resulting in the increase of the Gaussian-filtered 
value of the amplitude because of the larger value of the Gaussian filter 
function for a lower wavenumber. The negative-diffusion term represents this 
amplification of the filtered value resulting from the wavenumber shift. We 
introduce here a different interpretation of the negative diffusivity, using 
Fig.~\ref{fig2}. Let us consider the 1D pure advection of a step function,
\[
f(x,t=0)=\left\{ {{\begin{array}{*{20}l}
 1 \quad & {\mbox{for }x<x_0 ,} \\
 0 & {\mbox{otherwise},} \\
\end{array} }} \right.
\]
in a stretching velocity field $u=ax$, where $x_0 $ ($>0)$ is the initial 
position of the discontinuity in the step function and $a$ is a positive 
constant. The exact solution of $\partial f/\partial t+u\partial f/\partial 
x=0$ under this condition is
\[
f(x,t)=\left\{ {{\begin{array}{*{20}l}
 1 \quad & {\mbox{for }x<x_0 \exp (at),} \\
 0 & {\mbox{otherwise},} \\
\end{array} }} \right.
\]
which indicates that the discontinuous step will change its position without 
changing its height and profile; that is, the wavenumber shift does not 
occur in the unfiltered true solution of the present example. Applying the 
Gaussian filter to this solution smoothes out the discontinuity to yield a 
mollified step whose characteristic width is about $2\Delta $, where $\Delta 
$ is the characteristic length of the applied Gaussian filter. Obviously 
 the characteristic width of the mollified step is time-independent if 
$\Delta$ is constant. If, 
however, the pure advection equation in terms of the filtered value, 
$\partial \bar {f}/\partial t+u\partial \bar {f}/\partial x=0$, is used to 
advance the filtered profile (corresponding to the case where the residual 
stress term is clipped), the width of the mollified step, unlike that in the 
true solution, increases gradually as time goes by due to the stretching 
velocity field where a downstream fluid particle moves faster than an 
upstream particle. To counteract this artificial expansion of the mollified 
step, a modification by negative, not positive, diffusion is necessary, 
which sharpens $\bar {f}$. In the case of a compression velocity field, a 
similar but opposite treatment, i.e., the addition of a positive diffusion 
term, is needed because an artificial compression of the mollified step 
arises if only the pure advection equation is assumed.

The present physical picture may allow us to conclude that the subfilter-scale 
terms should have an analogous negative diffusivity also for any other 
filter functions that smooth the profile of dependent variables. In the 
above discussion, there is no reason that the filter shape must be Gaussian. 
The artificial expansion of the step discussed above must occur whenever the 
step is smoothed out by a smooth filter but the pure advection equation is 
solved. The above discussion also suggests that a numerically unstable term 
can appear by filtering even if the true solution (both filtered and 
unfiltered) is physically bounded.

\section{Towards a further generalization}
\label{secVI}
The present theory has several limitations in its applicability resulting 
from the assumptions and simplifications made. In this section we remark on 
some significant issues that must be resolved for further generalization.

To construct a more general theory for the numerical stability of the GFNS 
equations in statistically steady states, one has to elucidate the numerical 
stability of, not only Eq.~(\ref{eq28}), but also
\begin{equation}
\label{eq99}
\frac{\partial f_i }{\partial t}={\mathcal L}_{[ij]} f_j \quad \mbox{for }i=1,2,3,
\end{equation}
where ${\mathcal L}_{[ij]} $ ($i,j=1,2,3)$ are infinite sums of differential operators 
with position-dependent coefficients, determined by the expansion series, 
and in general
\[
{\mathcal L}_{[ab]} {\mathcal L}_{[cd]} \ne {\mathcal L}_{[cd]} {\mathcal L}_{[ab]} \quad \mbox{for }(a,b)\ne (c,d);
\]
that is, these operators do not commute with each other. Equation (\ref{eq99}) 
represents a complicated coupling between the equations of different 
velocity components. In the present study, having assumed 1D or 2D mean 
velocity fields, some of the mean-fluctuation terms were uncoupled as 
Eqs.~(\ref{eq31}), (\ref{eq32}), (\ref{eq61}), and (\ref{eq76}), and hence 
the determined stability conditions 
are accurate only for the corresponding uncoupled portions. In fully 3D 
cases where high-order velocity fields must be assumed, however, we may well 
have to treat the fully coupled system (\ref{eq99}).

The strongest assumption among those made in this paper may be the omission 
of the nonlinear fluctuation term ${\mathcal R}_{123} [u'_j \partial u'_i /\partial 
x_j ]$ in Eq.~(\ref{eq11}). We could not take into consideration the effects of this 
term since the theoretical investigation of it is quite difficult to perform 
accurately, and the present theory is thus only valid when the fluctuation 
is small. Terms of this type, as is well known, have a dissipative character 
in many turbulent flows, and hence when the dissipation of the omitted term 
is strong enough, the instability of the mean-fluctuation terms can be 
eliminated completely. One possible way to gain detailed knowledge of the 
nonlinear term is an {\it a priori} test using DNS, which enables us to determine the 
values of all terms in the GFNS equations. Observing and examining the 
numerically determined terms should allow us to obtain a more accurate 
prediction of the numerical instability. When, for example, the absolute 
value of the nonlinear term (plus the molecular viscosity) is smaller than 
that of the sum of the unstable terms, the instability can not be eliminated 
irrespective of the specific characteristic of the nonlinear term. Also, 
comparing the amounts of the energy dissipations due to the unstable terms 
and the nonlinear terms should provide a useful insight. This issue will be 
addressed in a future paper.

Lastly, we make a brief comment on cases with a nonuniform filter width. 
Consider again pure advection of a step function being smoothed by a smooth 
filter. When the filter width is spatially nonuniform, in the advection 
process the width of the mollified step varies according to its position 
even for a constant velocity, and hence a negative diffusion must take place 
at least in the period where the width decreases. Such an effect of nonuniform 
filtering must be discussed carefully in the near future.

\section{Conclusion}
\label{secVII}
We have presented a generalized theory for the numerical instability of the 
Gaussian-filtered Navier-Stokes equations. The theory allows for high-order 
mean velocity fields and high-order derivatives resulting from the Gaussian 
filtering operation. Also, we have described stability conditions regarding 
the choice of the filter widths in several situations, the violation of 
which should lead to unconditional numerical instability of the filtered 
system even when a completely accurate subfilter-scale model exists and is 
used. It is worth noting again that the closed formulas of the filtered 
mean-fluctuation terms determined under statistically steady-state 
conditions involve various kinds of unstable derivatives that, because their 
coefficients are time-independent, always exhibit numerical instability in a 
fixed range of directions. As has been proven by a simple example, the 
essential part of the present results can be true even if a non-Gaussian 
smooth filter is assumed.

We stress that if one skirts this numerical instability problem, the 
accuracy of the LES results will plateau. It is hard to imagine that ideally 
accurate solutions can be achieved by incorporating an artificial damping or 
clipping technique to avoid this numerical difficulty, because when the 
subfilter-scale terms act unstably, the absolute values of their unstable 
portions must be greater than that of the molecular and turbulent 
viscosities, and the adoption of such artificial techniques thus corresponds 
to the disregard of a term whose dominance is greater than that of a term 
involved {\it ab initio} in the Navier-Stokes equations. Recently, Moeleker and Leonard 
\cite{ref16} have tackled this numerical instability problem and proposed an 
approach to potentially resolve it, based on an anisotropic particle method 
incorporating a remeshing technique. Their method has provided excellent 
results for a 2D scalar advection-diffusion equation with a known velocity 
field. However, the extension of that approach to the 3D Navier-Stokes 
equations has to the author's knowledge not yet been achieved. Because 
finite difference schemes have been used widely in turbulence computations, 
constructing a stable and accurate solver in the finite difference framework 
would be preferable, though it will be an exceedingly difficult task and 
might even be an unsolvable problem, such as the gravitational three-body 
problem and the algebraic solution of general fifth-order polynomial 
equations of one variable. We do not know so far whether this instability 
problem is resolvable or not, but we can say that this problem is not 
something that can be avoided when an accurate solution is desired.

\begin{acknowledgments}
One of the authors (M.I.) thanks F.~Hamba for helpful and valuable 
discussions. Thanks are extended to A.~Yoshizawa and Y.~Morinishi for 
encouragement and comments. This work was supported by the Ministry of 
Education, Culture, Sports, Science, and Technology of Japan through the 
Grant-in-Aid for Young Scientists (B) (No. 17760151) and also under an IT 
research program ``Frontier Simulation Software for Industrial Science.''
\end{acknowledgments}

\appendix*

\section{Alternative derivation of the exact expansion series for 
Gaussian filters}
The exact expansion series for Gaussian filters has served as a powerful 
tool in our study. We present here a derivation of the series to improve the 
self-consistency of the present paper. This derivation seems to be rather 
intricate and drawn out compared to those by Moeleker and Leonard \cite{ref16} and 
by Carati et al.~\cite{ref15}, but it only consists of elementary mathematics: the 
Taylor expansion, integration by parts, and some simple algebraic 
operations. Some readers may prefer the present derivation.

Let $a(x)$, $b(x)$, $f(x)$, and $g(x)$ be arbitrary, differentiable and 
continuous functions of $x$. For the Gaussian filter with $\gamma =1/2$ and 
the characteristic width of $\Delta $, the exact expansion series in 1D 
reads
\begin{equation}
\label{eq100}
\overline {(ab)} =\sum\limits_{n=0}^\infty {\frac{\Delta 
^{2n}}{n!}\frac{\partial ^n\bar {a}}{\partial x^n}\frac{\partial ^n\bar 
{b}}{\partial x^n}} .
\end{equation}
In what follows, we derive the right-hand side of this equation from the 
left-hand side.

Taylor expanding $a$ with respect to $x$ results in
\begin{equation}
\label{eq101}
a=\sum\limits_{m=0}^\infty {A_m x^m} \quad \mbox{with}\quad A_m \equiv 
\frac{1}{m!}\left. {\frac{\partial ^ma}{\partial x^m}} \right|_{x=0} .
\end{equation}
Substituting this into $\overline {(ab)} $ yields
\begin{equation}
\label{eq102}
\overline {(ab)} =\overline {\left( {\sum\limits_{m=0}^\infty {A_m x^m} b} 
\right)} =\sum\limits_{m=0}^\infty {A_m \overline {(x^mb)} } .
\end{equation}
Successively using
\begin{equation}
\label{eq103}
\overline {(xf)} ={\mathcal L}\bar {f}\quad \mbox{with}\quad {\mathcal L}=x+\Delta 
^2\frac{\partial }{\partial x},
\end{equation}
which corresponds to Eq.~(\ref{eq15}) derived using integration by parts, we obtain 
the identity \cite{ref19}
\begin{equation}
\label{eq104}
\overline {(x^mf)} ={\mathcal L}^m\bar {f}.
\end{equation}
This rewrites Eq.~(\ref{eq102}) as
\begin{equation}
\label{eq105}
\overline {(ab)} =\sum\limits_{m=0}^\infty {[A_m ({\mathcal L}^m\bar {b})]} .
\end{equation}

The component ${\mathcal L}^m\bar {b}$ in Eq.~(\ref{eq105}) can further be rewritten as 
follows: Operating ${\mathcal L}$ once on $\bar {b}$ yields
\begin{eqnarray}
\label{eq106}
{\mathcal L}\bar {b} &=& x\bar {b}+\Delta ^2\frac{\partial }{\partial x}\bar {b} \nonumber \\
 &=& ({\mathcal L}^1\cdot 1)\left( {\Delta ^2\frac{\partial }{\partial x}} \right)^0\bar 
{b}+({\mathcal L}^0\cdot 1)\left( {\Delta ^2\frac{\partial }{\partial x}} \right)^1\bar 
{b},
\end{eqnarray}
where
\begin{equation}
\label{eq107}
{\mathcal L}^1\cdot 1\equiv \left( {x+\Delta ^2\frac{\partial }{\partial x}} 
\right)1=x\quad \mbox{and}\quad {\mathcal L}^0\cdot 1=1.
\end{equation}
We introduce here
\begin{equation}
\label{eq108}
B_{N,M} \equiv ({\mathcal L}^N\cdot 1)\left( {\Delta ^2\frac{\partial }{\partial x}} 
\right)^M\bar {b}.
\end{equation}
Based on Eq.~(\ref{eq106}), definition (\ref{eq108}), and
\[
{\mathcal L}(gf)=({\mathcal L}g)f+g\left( {\Delta ^2\frac{\partial }{\partial x}} \right)f,
\]
the following identities are derived:
\begin{eqnarray}
\label{eq109}
{\mathcal L}\bar {b} &=& {\mathcal L}B_{0,0} \nonumber \\ 
 &=& B_{1,0} +B_{0,1} , 
\end{eqnarray}
\begin{eqnarray}
\label{eq110}
{\mathcal L}B_{N,M} &=& ({\mathcal L}^{N+1}\cdot 1)\left( {\Delta ^2\frac{\partial }{\partial x}} 
\right)^M\bar {b}+({\mathcal L}^N\cdot 1)\left( {\Delta ^2\frac{\partial }{\partial x}} 
\right)^{M+1}\bar {b} \nonumber \\ 
 &=& B_{N+1,M} +B_{N,M+1} ,
\end{eqnarray}
which allow us to obtain
\begin{eqnarray}
\label{eq111}
{\mathcal L}^2\bar {b}&=&{\mathcal L}^2B_{0,0} ={\mathcal L}B_{1,0} +{\mathcal L}B_{0,1} \nonumber \\ 
 &=& B_{2,0} +2B_{1,1} +B_{0,2} , \nonumber \\ 
 {\mathcal L}^3\bar {b} &=& {\mathcal L}^3B_{0,0} ={\mathcal L}B_{2,0} +2{\mathcal L}B_{1,1} +{\mathcal L}B_{0,2} \nonumber \\ 
 &=& B_{3,0} +3B_{2,1} +3B_{1,2} +B_{0,3} , \\ 
 &&\cdots \nonumber \\ 
{\mathcal L}^m\bar {b} &=& \sum\limits_{n=0}^m {s_{m-n,n} B_{m-n,n} } . \nonumber
\end{eqnarray}
Here, the coefficients $s_{m-n,n} $ ($m=0,1,2\ldots $ and $n=0,1,\ldots ,m)$ 
form a so-called Pascal's triangle, and thus
\begin{equation}
\label{eq112}
s_{m-n,n} =\frac{m!}{n!(m-n)!}.
\end{equation}
From Eqs.~(\ref{eq108}), (\ref{eq111}), and (\ref{eq112}), we have
\begin{equation}
\label{eq113}
{\mathcal L}^m\bar {b}=\sum\limits_{n=0}^m {\frac{\Delta 
^{2n}}{n!}\frac{m!({\mathcal L}^{m-n}\cdot 1)}{(m-n)!}\frac{\partial ^n\bar 
{b}}{\partial x^n}} .
\end{equation}

Substituting Eq.~(\ref{eq113}) into Eq.~(\ref{eq105}) yields
\begin{equation}
\label{eq114}
\overline {(ab)} =\sum\limits_{m=0}^\infty {\left[ {A_m \sum\limits_{n=0}^m 
{\frac{\Delta ^{2n}}{n!}\frac{m!({\mathcal L}^{m-n}\cdot 1)}{(m-n)!}\frac{\partial 
^n\bar {b}}{\partial x^n}} } \right]} .
\end{equation}
Using Eq.~(\ref{eq104}), $({\mathcal L}^{m-n}\cdot 1)$ in this equation can easily be rewritten 
into $\overline {(x^{m-n})} $, which can further be rewritten as
\begin{equation}
\label{eq115}
\overline {(x^{m-n})} =\frac{(m-n)!}{m!}\overline {\left( {\frac{\partial 
^nx^m}{\partial x^n}} \right)} .
\end{equation}
Substituting this into Eq.~(\ref{eq114}) yields
\begin{equation}
\label{eq116}
\overline {(ab)} =\sum\limits_{m=0}^\infty {\left[ {A_m \sum\limits_{n=0}^m 
{\frac{\Delta ^{2n}}{n!}\overline {\left( {\frac{\partial ^nx^m}{\partial 
x^n}} \right)} \frac{\partial ^n\bar {b}}{\partial x^n}} } \right]} .
\end{equation}
Moreover, because
\begin{equation}
\label{eq117}
\frac{\partial ^nx^m}{\partial x^n}=0\quad \mbox{for }n>m,
\end{equation}
the summation over $n=1,2,\ldots ,m$ in Eq.~(\ref{eq116}) can be extended to that 
over $n=1,2,\ldots ,\infty $ to obtain
\begin{equation}
\label{eq118}
\overline {(ab)} =\sum\limits_{m=0}^\infty {\left[ {\sum\limits_{n=0}^\infty 
{\frac{\Delta ^{2n}}{n!}A_m \overline {\left( {\frac{\partial 
^nx^m}{\partial x^n}} \right)} \frac{\partial ^n\bar {b}}{\partial x^n}} } 
\right]} .
\end{equation}
Based on the commutativity between differentiations and filtering, after 
some mathematical operations we finally obtain Eq.~(\ref{eq100}).

\newpage 

\newpage 

\end{document}